\documentclass{amsart}

\newtheorem{theorem}{Theorem}[section]
\newtheorem{lemma}[theorem]{Lemma}

\theoremstyle{definition}

\newtheorem{xca}[theorem]{Exercise}

\theoremstyle{remark}

\numberwithin{equation}{section}

\begin{document}

\title{Finding Paths and Cycles in Graphs}

\author{Sergey Gubin}
\address{Genesys Telecommunication Laboratories, Inc.
\newline\indent
1255 Treat Blvd., Walnut Creek, CA 94596
}
\email{sgubin@genesyslab.com}


\subjclass[2000]{Primary 05C38, Secondary 68R10}

\date{August 30, 2007}


\keywords{graph, path, cycle, Hamiltonian, clique, NP-complete}

\begin{abstract}
A polynomial time algorithm which detects all paths and cycles of all lengths in form of vertex pairs (start, finish).
\end{abstract}

\maketitle
\section*{Introduction}

Problem of path/cycle existence and finding is a big topic in Discrete Mathematics and Computer Science \cite[and others]{Hamilton1, Hamilton2, Tutte1, Tutte2, Ore, Karp, Chvatal, Jung, Garey, Fan, Yann, Bauer1, Diestel, TSP, Bauer2, Cook}. This article is not a review but just another design.
\newline\indent
The design's idea may be traced back to Sir William R. Hamilton, \cite{Hamilton1}: ``I have lately been led to the conception of a new system, or rather family of systems, of non-commutative roots of unity, ... which admit, even more easily than the quaternion symbols do, of geometrical interpretation. ... every one of the results may be interpreted, as having reference to the passage from face to face (or from corner to corner) of the icosahedron (or of the dodecahedron) ... .'' Let us rephrase: a formal system was what allowed tracking of the ``passages''.
\newline\indent
This work models a graph's walks with a generative grammar and applies dynamic programming to detect paths/cycles. Some deviation from the widely accepted terminology is due to the grammar: walks are just sequences of adjacent arcs, arcs/loops are walks/paths/circuits/cycles of length $1$, etc. Author hopes that this will not be a nuisance. Also, any graph becomes a multi graph when talking walks. So, let us start with that from the very beginning.

\section{Walk generator}
\label{s:wg}

Let $g = (V, A)$ be a given multi digraph (possible with loops): $V$ is the vertex set and $A$ is the arc set of the graph.
Let's arbitrarily enumerate vertices:
\[
V = \{v_{1}, v_{2}, \ldots, v_{n} \},
\]
- and build a set matrix $S=(s_{ij})_{n\times n}$ whose $(i,j)$-element is the set of all arcs which start in vertex $v_{i}$ and finish in vertex $v_{j}$; some of the elements can be $\emptyset$.
\newline\indent
Let's define powers of matrix $S$:
\[
S^{1} = S, ~ S^{k+1} = S^{k} S, ~ k \geq 1.
\]
The multiplication of set matrices is performed by the rules of the usual matrix multiplication but with replacement of the numeric products with Cartesian products of sets and the numerical sum with set joins: if $X=(x_{ij})_{a\times b}$ and $Y=(y_{ij})_{b\times c}$ are set matrices, then
\[
X Y = ( \bigcup_{\beta = 1}^{b} x_{i\beta} \times y_{\beta j} )_{a \times c}.
\]
Let us mention that the set matrix multiplication is an associative operation. For matrix $S^{k}$, the last formula gives:
\begin{equation}
\label{e:wg}
S^{1} = S, ~ S^{k+1} = ( \bigcup_{\beta = 1}^{n} (S^{k})_{i\beta} \times (S)_{\beta j} )_{n \times n}, ~ k \geq 1.
\end{equation}
Here and further, symbol $(M)_{ij}$ means $(i,j)$-element of matrix $M$.
\newline\indent
Matrix $S$ is a walk generator: element $(S^{k})_{ij}$ consists of all such walks ``written'' in arcs, which have length $k$, start in vertex $v_{i}$, and finish in vertex $v_{j}$; all circuits of length $k$ are assembled in diagonal elements. That may be cleared with the following decomposition:
\[(S^{k+1})_{ij} = \bigcup_{\beta = 1}^{n} (S^{k})_{i\beta} \times (S)_{\beta j} =
\]
\[
\bigcup_{\beta = 1}^{n} ( \bigcup_{\gamma=1}^{n}(S^{k-1})_{i\gamma} \times (S)_{\gamma\beta}) \times (S)_{\beta j} = \bigcup_{\beta, \gamma = 1}^{n} (S^{k-1})_{i\gamma} \times (S)_{\gamma\beta} \times (S)_{\beta j} = \ldots =
\]
\begin{equation}
\label{e:wg-factorial}
= \bigcup_{\sigma} (S)_{\sigma_{1}=i,\sigma_{2}} \times (S)_{\sigma_{2}\sigma_{3}} \times \ldots (S)_{\sigma_{k-1}\sigma_{k}} \times (S)_{\sigma_{k},\sigma_{k+1}=j},
\end{equation}
- where join is taken over all $n^{k-1}$ samples $\sigma = (\sigma_{1}, \sigma_{2}, \ldots, \sigma_{k+1})$ of $k+1$ numbers constrained with the following:
\begin{equation}
\label{e:wg-constrain}
\sigma_{1} = i; ~ 1 \leq \sigma_{m} \leq n, ~ m = 2,3,...,k; ~ \sigma_{k+1} = j.
\end{equation}
Let $(S_{P}^{k+1})_{ij}$ be a limitation of join \ref{e:wg-factorial} to those samples $\sigma$ which are permutations; and $(S_{C}^{k+1})_{ii}$ be a limitation of join \ref{e:wg-factorial} to those samples $\sigma$ which are circular permutations. 
\begin{theorem}
\label{t:wg}
In multi digraph $g$:
\newline
1. There is a walk of length $k > 1$ from $v_{i}$ into $v_{j}$ iff $(S^{k})_{ij} \neq \emptyset$;
\newline
2. There is a path of length $k > 1$ from $v_{i}$ into $v_{j}$ iff $(S_{P}^{k})_{ij} \neq \emptyset$;
\newline
3. There is a cycle of length $k > 1$ attached to $v_{i}$ iff $(S_{C}^{k})_{ii} \neq \emptyset$;
\newline
4. There are paths and cycles of length $1$ iff appropriate elements of walk generator $S = S^{1}$ are not empty.
\end{theorem}

\begin{xca}
\label{x:dion}
Suppose $g$ has two vertices $v_{1}, ~ v_{2}$ and four arcs $a_{11},~a_{12},~a_{21}$, and $a_{22}$: arc $a_{ij}$ begins in vertex $v_{i}$ and ends in vertex $v_{j}$. Then, the walk generator for $g$ is the following set matrix:
\[
S^{1} = S = \left (
\begin{array}{ll}
\{a_{11}\}&\{a_{12}\} \\
\{a_{21}\}&\{a_{22}\} \\
\end{array} \right ).
\]
The arcs/loops are walks of length $1$. Arcs $a_{12}$ and $a_{21}$ are paths of length $1$. Loops $a_{11}$ and $a_{22}$ are cycles of length $1$. 
\newline\indent
Walks of length $2$ constitute elements of set matrix $S^{2}$:
\[
S^{2} = S S = \left (
\begin{array}{ll}
\{a_{11}a_{11},a_{12}a_{21}\}&\{a_{11}a_{12},a_{12}a_{22}\} \\
\{a_{21}a_{11},a_{22}a_{21}\}&\{a_{21}a_{12},a_{22}a_{22}\} \\
\end{array} \right ).
\]
There are no paths of length $2$. Diagonal elements are circuits of length $2$. Among them is cycle $(a_{12}a_{21}) = (a_{21}a_{12})$. It is a Hamiltonian cycle.
\newline\indent
Walks of length $3$ constitute elements of set matrix $S^{3} = S^{2} S$, and so on.
\end{xca}
Iterations \ref{e:wg} create a language - we may call it an arc language. Its alphabet is $A$, and its grammar is a context-sensitive L-system whose axiom is walk generator $S$ and whose production rules, in essence, consist of omitting the curly braces in formula \ref{e:wg} and replacing $\emptyset$ with empty string. The $k$-th generation of the system consists of all walks/strings of length $k$ written in arcs. The walks are sorted over vertex pairs (start, finish). 
\newline\indent
The exact numbers of $k$-walks in these vertex pairs (start, finish) are elements of $k$-th power of the multi digraph's adjacency matrix. From formula \ref{e:wg-factorial}, a rough upper bound for the numbers is 
\[
|V|^{k-1}(\max_{ij}|S_{ij}|)^{k}.
\]
The bound is reachable in complete graphs. General walk pattern is a path with cycles attached, and multiplicand $|V|^{k-1}$ is  due to the branching. Nevertheless, using formula \ref{e:wg} can be an efficient method for particular problems/graphs - sparse digraphs, trees, and others with simple walk pattern.
\begin{xca}
\label{x:dion1}
In the example from exercise \ref{x:dion}, there are $2^{k+1}$ walks of length $k, ~ k \geq 1$, because there is cycle $a_{12}a_{21}$ with two cycles/loops attached to it - $a_{11}$ and $a_{22}$. Written in arcs, the pattern of walks from $v_{i}$ into $v_{j}$ is a self-similar structure described by the following grammar:
\[
[~[a_{ii}]a_{ij}[a_{jj}]a_{ji}~]~a_{ij}~[~[a_{jj}]a_{ji}[a_{ii}]a_{ij}~],
\]
- where symbol $[x]$ means zero or more repetitions of x.
\end{xca}
For the path/cycle problem, the algorithmic simplicity of formula \ref{e:wg} looks tempting: all paths/cycles could be found with filtering of all $n^{2}$ sets $(S^{k})_{ij}$. Although those sets have exponential power, filtering per se could be made efficient due to a possibility of sorting/filtering the sets during iterations \ref{e:wg}. That which makes the method inefficient is a need for calculation/remembering of exponential size sets - elements of $S^{k}$. The problem can be tried and fought with walk coloring.

\section{Walk coloring}
\label{s:ac}

Walk colors are a semantics of the arc language defined by formula \ref{e:wg}. Coloring can be done with an arc coloring; a function defined on walks; and with other means, as well. For example: adjacency matrix is a coloring with a ``number of walks'' function; transitional diagram is a digraph arc-colored with the events triggering transitions; Markov chain is a digraph arc-colored with the transitional probabilities; the Icosian Calculus \cite{Hamilton1, Hamilton2} is the icosahedron colored with the Hamilton's symbols, etc. The original walk generator defined in section \ref{s:wg} is a coloring appropriate to the identity function on set of all walks. Its opposite is a monochromatic coloring - all walks have the same color. 
\begin{xca}
\label{x:coin}
In the example from exercise \ref{x:dion}, let's color the arcs starting in $v_{1}$ with color ``Head'' and the arcs starting in $v_{2}$ with color ``Tail''. Also, let's use the string presentation of iterations \ref{e:wg} described in that section. Then, the example will become an algorithmic model of coin flipping:
\[
S = \left (
\begin{array}{ll}
H&H \\
T&T \\
\end{array} \right ), ~
S^{2} = \left (
\begin{array}{ll}
HH,HT&HH,HT \\
TH,TT&TH,TT \\
\end{array} \right ), ~ \ldots
\]
Changing of the colors to ``$0$'' and ``$1$'' will make the example a generator of all/random finite $(0,1)$-sequences.
\end{xca}
\begin{xca}
\label{x:hamilton}
Let $P$ be the following monochromatic arc coloring of $g$:
\[
(P)_{ij} = \left \{ \begin{array}{ll}
1, & (S)_{ij} \neq \emptyset \\
0, & (S)_{ij} = \emptyset \\
\end{array}. \right.
\]
Then, number of different Hamiltonian cycles in $g$ is
\[
\frac{1}{n}\sum_{\sigma} (P)_{\sigma_{1}\sigma_{2}}  (P)_{\sigma_{2}\sigma_{3}}  \ldots  (P)_{\sigma_{n-1} \sigma_{n}}  (P)_{\sigma_{n} \sigma_{1}},
\]
where sum is taken over all circular permutations $\sigma = (\sigma_{1},\sigma_{2},\ldots,\sigma_{n})$ of numbers $1,2,\ldots,n$.
\end{xca}

In the monochromatic case, color is just a sign of adjacency. The coloring can be modeled with Boolean matrix - a matrix filled with $true$ and $false$ (for $\emptyset$). Then, iterations \ref{e:wg} can be replaced with formulas for powers of that Boolean matrix. The powers are calculated with the algorithm for numeric matrix multiplication except the products are replaced with conjunctions and the sums are replaced with disjunctions: if $X=(x_{ij})_{a\times b}$ and $Y=(y_{ij})_{b\times c}$ are Boolean matrices, then
\begin{equation}
\label{e:b0}
X Y = ( \bigvee_{\beta = 1}^{b} x_{i\beta} \wedge y_{\beta j} )_{a \times c}.
\end{equation}
Let us mention that the Boolean matrix multiplication is an associative operation. 
\newline\indent
Boolean arc coloring is as follows:
\begin{equation}
\label{e:ba}
(B^{1})_{ij} = \left \{ \begin{array}{ll}
true, & (S)_{ij} \neq \emptyset \\
false, & (S)_{ij} = \emptyset \\
\end{array} \right.
\end{equation}
And Boolean walk coloring is defined with the following iterations:
\begin{equation}
\label{e:bg}
B^{k+1} = ( \bigvee_{\beta = 1}^{n} (B^{k})_{i\beta} \wedge (B^{1})_{\beta j} )_{n \times n}, ~ k \geq 1.
\end{equation}
The $true/false$ coloring is efficient for the reachability problems: formula \ref{e:bg} finds all pairs of connected vertices with $(2n-1)n^{3}$ operations ``$\vee$'' and ``$\wedge$'' at most.
\newline\indent
To prove that, let's perform decomposition of $(B^{k})_{ij}$ exactly like for formula \ref{e:wg-factorial}:
\begin{equation}
\label{e:dec}
(B^{k+1})_{ij} = \bigvee_{\sigma} (B^{1})_{\sigma_{1}=i,\sigma_{2}} \wedge (B^{1})_{\sigma_{2}\sigma_{3}} \wedge \ldots (B^{1})_{\sigma_{k-1}\sigma_{k}} \wedge (B^{1})_{\sigma_{k},\sigma_{k+1}=j},
\end{equation}
- where disjunction is taken over all number samples $\sigma$ which satisfy constrains \ref{e:wg-constrain}. 
\newline\indent
Let $(B_{P}^{k+1})_{ij}$ be a limitation of disjunction \ref{e:dec} to those samples $\sigma$ which are permutations; and $(B_{C}^{k+1})_{ii}$ be a limitation of disjunction \ref{e:dec} to those samples $\sigma$ which are circular permutations. Then, theorem \ref{t:wg} implies
\begin{theorem}
\label{t:bc}
In multi digraph $g$:
\newline
1. There is a walk of length $k > 1$ from $v_{i}$ into $v_{j}$ iff $(B^{k})_{ij} = true$;
\newline
2. There is a path of length $k > 1$ from $v_{i}$ into $v_{j}$ iff $(B_{P}^{k})_{ij} = true$;
\newline
3. There is a cycle of length $k > 1$ attached to $v_{i}$ iff $(B_{C}^{k})_{ii} = true$;
\newline
4. There are paths and cycles of length $1$ iff appropriate elements of Boolean matrix $B^{1}$ are $true$.
\end{theorem}

Although Boolean coloring \ref{e:ba}, \ref{e:bg} is inefficient for the path/cycle problem in multi digraph, it still can be efficient for other particular problems.
\begin{xca}[Shortest Path]
\label{x:shortest path}
Length of the shortest path from vertex $v_{i_{0}}$ into vertex $v_{j_{0}}$ can be calculated with formula \ref{e:bg} by multiplication of $i_{0}$-th string of Boolean matrix $B^{k}$ on whole Boolean matrix $B^{1}$ until element $(B^{k+1})_{i_{0}j_{0}}$ will become $true$, $k = 1,2,\ldots,n-2$. Then, the shortest path has length $k+1$. The table below compares the method with classical routing algorithms:
\[
\begin{array}{r|ccc}
Algorithm & \mbox{Formula \ref{e:bg}} & \mbox{Bellman-Ford} & \mbox{Dijkstra} \\
\hline
Time~complexity & O(|V|^{3})& O(|V||A|)& O(|V|^{2}+|A|)\\
\end{array}
\]
Let us emphasize the algorithmic simplicity of formula \ref{e:bg} and the potential of that formula for further computational simplifications.
\end{xca}
\begin{xca}[Maximum Clique]
\label{x:clique}
Let $Q$ be the following Boolean matrix:
\[
(Q)_{ij} = \left \{ \begin{array}{cl}
(B^{1})_{ij} \wedge (B^{1})_{ji}, & i > j \\
false, & i \leq j \\
\end{array}, \right.
\]
- where $B^{1}$ is defined with formula \ref{e:ba}. Let's define $Q^{m}$ with the following iterations:
\[
Q^{1} = Q, ~ Q^{m+1} = Q^{m}Q \wedge Q^{m}, ~ m \geq 1
\]
- where product of Boolean matrices is defined with formula \ref{e:b0}, and conjunction of Boolean matrices is a matrix of conjunctions of appropriate elements. It may be seen that maximum clique in $g$ has the following size:
\[
q = \min_{k} \{k ~|~ Q^{k} = (false)_{n\times n}\}.
\]
Let us clarify that any vertex is a clique of size $1$ and sketch a proof:
\begin{proof}[Sketch of proof]
Boolean matrix $Q$ defines a subgraph of multi digraph $g$. This subgraph is a directed forest. All walks in this forest are paths. Boolean matrix $Q^{k}$ rids that forest's arc set of all arcs except those vertex pairs $(v_{i},v_{j})$ which are start and finish of paths of all length $1,2,\ldots,k$. Cliques of size $k+1$ are the only invariants under such depletion of $g$.
\end{proof}
\end{xca}

For multi digraph, the path/cycle problem's semantics is ``to visit vertices only once''. That gives an idea of efficient walk coloring for the problem. The idea is to make from the walk colors a memory. It is an interesting question, what to remember. Specter of options stretches from remembering visited vertices only to remembering unvisited vertices only. Disconnected graphs make the strategy question non-trivial. Nevertheless, those extreme strategies are dual and both based on an innate knowledge \ref{e:wg}. Let's explore the rationalistic approach.

\section{Path problem}
\label{s:dc}

Let's encode walks with the unvisited vertices. Two opposite strategies are possible: to remember arcs' start-vertices or to remember arcs' finish-vertices. The strategies are dual. They correspond to the ambiguity of words ``visited vertex'': was it exited already or just entered?
\newline\indent
Arc coloring/coding with start-vertices is as follows:
\begin{equation}
\label{e:dc-sa}
(F^{1})_{ij} = \left \{
\begin{array}{cl}
V - \{v_{i}\}, & (S)_{ij} \neq \emptyset ~ \wedge ~ i \neq j \\
\emptyset, & (S)_{ij} = \emptyset ~ \vee ~ i = j \\
\end{array},
\right.
\end{equation}
- where $S$ the walk generator. Walk coloring with start-vertices is defined with the following iterations:
\begin{equation}
\label{e:dc-sw} 
(F^{k+1})_{ij} = \left \{ \begin{array}{cl}
\bigcup_{\beta=1}^{n} (F^{k})_{i\beta} \cap (F^{1})_{\beta j},& i \neq j \\
\emptyset, & i = j \\
\end{array}, \right. ~ k \geq 1
\end{equation}
The formula may be seen as another set matrix multiplication. 
\newline\indent
Let's call that coloring defined with formulas \ref{e:dc-sa} and \ref{e:dc-sw} a $F$-coloring: all arcs which start from vertex $v_{i}$ have the same color; the color is set $V - \{v_{i}\}$; the diagonal elements are rid of because they are loops/circuits; all walks from vertex $v_{i}$ into vertex $v_{j}$ have the same color; the color is set of all vertices which were missed by one or more walks from $v_{i}$ into $v_{j}$ - those vertices missed by a walk will extend that walk in the future when grammar \ref{e:wg} will allow that. To clarify that, let's perform decomposition of a non-empty set $(F^{k+1})_{ij}$:
\[
(F^{k+1})_{ij} = \bigcup_{\beta=1}^{n} (F^{k})_{i\beta} \cap (F^{1})_{\beta j} = 
\]
\[
= \bigcup_{\beta=1}^{n} \{ \bigcup_{\gamma = 1}^{n}(F^{k-1})_{i\gamma} \cap (F^{1})_{\gamma\beta} \} \cap (F^{1})_{\beta j} 
= \bigcup_{\beta,\gamma=1}^{n} (F^{k-1})_{i\gamma} \cap (F^{1})_{\gamma\beta}  \cap (F^{1})_{\beta j} = \ldots = 
\]
\begin{equation}
\label{e:dc-sw-decomposition}
= \bigcup_{\sigma} (F^{1})_{\sigma_{1}=i,\sigma_{2}} \cap (F^{1})_{\sigma_{2}\sigma_{3}} \cap \ldots \cap (F^{1})_{\sigma_{k-1} \sigma_{k}} \cap (F^{1})_{\sigma_{k}, \sigma_{k+1}=j},
\end{equation}
- where join is taken over all number samples $\sigma = (\sigma_{1},\sigma_{2},\ldots,\sigma_{k+1})$ which satisfy constrains \ref{e:wg-constrain}.
\newline\indent
That which makes $F$-coloring special is the following inclusion:
\begin{equation}
\label{e:dc-v}
(F^{k})_{ij} \subseteq V.
\end{equation}
It will be shown that this inclusion makes the coloring an efficient one. 

\begin{xca}
\label{x:dc-sa}
For the example from exercise \ref{x:dion}:
\[
F^{1} = \left (
\begin{array}{cc}
\emptyset&\{v_{2}\} \\
\{v_{1}\}&\emptyset \\
\end{array} \right ),
\]
\[
F^{2} = \left (
\begin{array}{cc}
\emptyset & \emptyset \cap \{v_{2}\} ~ \cup ~ \{v_{2}\} \cap \emptyset  \\
\{v_{1}\} \cap \emptyset ~ \cup ~ \emptyset \cap \{v_{1}\} & \emptyset \\
\end{array} \right ) = \left (
\begin{array}{ll}
\emptyset&\emptyset \\
\emptyset&\emptyset \\
\end{array} \right ).
\]
Thus, due to formula \ref{e:dc-sw}, $F^{k} = (\emptyset)_{2 \times 2}$ for all powers $k \geq 2$.
\end{xca}

\begin{lemma}
\label{t:dc-s}
In multi digraph $g$, if there is a path of length $k \geq 1$ from vertex $v_{i}$ into vertex $v_{j}$, then $(F^{k})_{ij} \neq \emptyset$.
\end{lemma}
\begin{proof}
Suppose, vertices
\[
v_{\mu_{1} = i},v_{\mu_{2}},\ldots,v_{\mu_{k}},v_{\mu_{k+1} = j}
\]
constitute a path of length $k$ from $v_{i}$ into $v_{j}$. Then, 
\[
v_{\mu_{k+1}} \neq v_{\mu_{m}}, ~ m = 1,2,\ldots,k.
\]
Thus, by definition \ref{e:dc-sa}:
\[
v_{\mu_{k+1}} \in (F^{1})_{\mu_{m}\mu_{m+1}} \neq \emptyset, ~ m = 1,2,\ldots,k.
\]
Then, due to decomposition \ref{e:dc-sw-decomposition}:
\[
v_{\mu_{k+1}} \in (F^{1})_{\mu_{1}\mu_{2}} \cap (F^{1})_{\mu_{2}\mu_{3}} \cap \ldots \cap (F^{1})_{\mu_{k}\mu_{k+1}} \subseteq (F^{k})_{\mu_{1}\mu_{k+1}} = (F^{k})_{ij} \neq \emptyset.
\]
\end{proof}
\begin{lemma}
\label{t:dc-s3}
In multi digraph $g$, if $(F^{k})_{ij} \neq \emptyset$, then there is a path of length $k$ from vertex $v_{i}$ into vertex $v_{j}$.
\end{lemma}
\begin{proof}
Let's use mathematical induction over $k$. 
\newline\indent
For k = 1, by definition \ref{e:dc-sa}, inequality $(F^{1})_{ij} \neq \emptyset$ implies
\[
(S)_{ij} \neq \emptyset.
\]
Then, there is an ark $a \in (S)_{ij}$. The arc is a path of length $1$. 
\newline\indent
Let's assume that the lemma is true for $k=l$.
\newline\indent
Let $k = l+1$. Due to presentation \ref{e:dc-sw-decomposition}, inequality $(F^{k})_{ij} \neq \emptyset$ implies existence of such number sample $\sigma_{0} = (\mu_{2},\mu_{3},\ldots,\mu_{k})$ that
\[
F_{\sigma_{0}} = (F^{1})_{i=\mu_{1},\mu_{2}} \cap (F^{1})_{\mu_{2}\mu_{3}} \cap (F^{1})_{\mu_{3}\mu_{4}} \cap \ldots \cap (F^{1})_{\mu_{k-1},\mu_{k}} \cap (F^{1})_{\mu_{k},\mu_{k+1} = j} \neq \emptyset.
\]
Then, due to definition \ref{e:dc-sa}, there are $k$ arcs
\[
a_{\mu_{m}\mu_{m+1}} \in (S)_{\mu_{m}\mu_{m+1}}, ~ m = 1,2,\ldots,k
\]
- where $S$ is the walk generator of $g$. The arcs create a walk $w$ of length $k$ from $v_{i}$ into $v_{j}$:
\[
w = (v_{\mu_{1}=i},a_{\mu_{1}\mu_{2}},v_{\mu_{2}},a_{\mu_{2}\mu_{3}},\ldots,v_{\mu_{k}},a_{\mu_{k}\mu_{k+1}}v_{\mu_{k+1}=j}).
\]
Let's notice that 
\[
(F^{k-1})_{\mu_{1}\mu_{k}} \supseteq (F^{1})_{i=\mu_{1},\mu_{2}} \cap (F^{1})_{\mu_{2}\mu_{3}} \cap (F^{1})_{\mu_{3}\mu_{4}} \cap \ldots \cap (F^{1})_{\mu_{k-1},\mu_{k}} \supseteq F_{\sigma_{0}} \neq \emptyset.
\]
Thus, due to the induction hypothesis, the following sub-walk of walk $w$
\[
(v_{\mu_{1}=i},a_{\mu_{1}\mu_{2}},v_{\mu_{2}},a_{\mu_{2}\mu_{3}},\ldots,v_{\mu_{k}})
\]
is a path of length $k-1$ from $v_{i=\mu_{1}}$ into $v_{\mu_{k}}$. From the other side, 
\[
(F^{k-1})_{\mu_{2}\mu_{k+1}} \supseteq (F^{1})_{\mu_{2}\mu_{3}} \cap (F^{1})_{\mu_{3}\mu_{4}} \cap \ldots \cap (F^{1})_{\mu_{k-1},\mu_{k}}\cap (F^{1})_{\mu_{k},\mu_{k+1} = j} \supseteq F_{\sigma_{0}} \neq \emptyset,
\]
as well. Thus, due to the induction hypothesis, the following sub-walk of walk $w$
\[
(v_{\mu_{2}},a_{\mu_{2}\mu_{3}},\ldots,v_{\mu_{k}},a_{\mu_{k}\mu_{k+1}}v_{\mu_{k+1}=j})
\]
is a path of length $k-1$ from $v_{\mu_{2}}$ into $v_{\mu_{k+1}=j}$. Thus, if
\[
v_{i} \neq v_{j},
\]
then walk $w$ will be a path of length $k$ from $v_{i}$ into $v_{j}$. That last inequality follows from definition \ref{e:dc-sw}.
\end{proof}
Let us emphasize the role of grammar \ref{e:wg} on an example of a disconnected graph.
\begin{xca}
\label{x:dm-x2}
Let's add to the digraph from exercise \ref{x:dion} vertex $v_{3}$, arc $a_{23}$ from $v_{2}$ into $v_{3}$, and isolated vertex $v_{4}$. Walk generator for the resulting graph is
\[
S = \left (
\begin{array}{cccc}
\{a_{11}\}&\{a_{12}\}&\emptyset&\emptyset \\
\{a_{21}\}&\{a_{22}\}&\{a_{23}\}&\emptyset \\
\emptyset&\emptyset&\emptyset&\emptyset \\
\emptyset&\emptyset&\emptyset&\emptyset \\
\end{array} \right ).
\]
There are three paths of length $1$ in the digraph: $(v_{1}a_{12}v_{2})$, $(v_{2}a_{21}v_{1})$, and $(v_{2}a_{23}v_{3})$. Also, there is one path of length $2$ in the digraph: $(v_{1}a_{12}v_{2}a_{23}v_{3})$.
\newline\indent
Arc $F$-coloring \ref{e:dc-sa} detects the $1$-paths:
\[
F^{1} = \left (
\begin{array}{cccc}
\emptyset&\{v_{2},v_{3},v_{4}\}&\emptyset&\emptyset \\
\{v_{1},v_{3},v_{4}\}&\emptyset&\{v_{1},v_{3},v_{4}\}&\emptyset \\
\emptyset&\emptyset&\emptyset&\emptyset \\
\emptyset&\emptyset&\emptyset&\emptyset \\
\end{array} \right ).
\]
Iterations \ref{e:dc-sw} detect the $2$-path and absence of the longer paths:
\[
F^{2} = \left (
\begin{array}{cccc}
\emptyset&\emptyset&\{v_{3},v_{4}\}&\emptyset \\
\emptyset&\emptyset&\emptyset&\emptyset \\
\emptyset&\emptyset&\emptyset&\emptyset \\
\emptyset&\emptyset&\emptyset&\emptyset \\
\end{array} \right ); ~
F^{k} = (\emptyset)_{4 \times 4}, ~ k \geq 3.
\]
\end{xca}
Together, lemmas \ref{t:dc-s} and \ref{t:dc-s3} constitute a theorem:
\begin{theorem}
\label{t:dc-s0}
In multi digraph $g$, there is a path of length $k \geq 1$ from vertex $v_{i}$ into vertex $v_{j}$ iff 
$(F^{k})_{ij} \neq \emptyset$.
\end{theorem}
Let's estimate computational complexity of theorem \ref{t:dc-s0}. Due to inclusion \ref{e:dc-v}, arc $F$-coloring \ref{e:dc-sa} requires $O(n^{3})$ operations. And walk $F$-coloring \ref{e:dc-sw} requires $O(kn^{4})$ operations: $O(n^{2})$ operations at most to calculate each of $n^{2}$ elements $(F^{m})_{ij }, ~ m = 1,2,\ldots,k$. The total computational complexity of theorem \ref{t:dc-s0} is 
\[
O(kn^{4}).
\]
Particularly, when $k = n-1$, theorem \ref{t:dc-s0} solves the Hamilton path problem in time
\[
O(n^{5}).
\]
\indent
Obviously, the results are true when an ``arcs' finish-vertices'' coloring is used instead of the ``arcs' start-vertices'' coloring described. A finish-vertices coloring can be done with the following $G$-coloring. Arc $G$-coloring is defined as follows:
\begin{equation}
\label{e:dc-fa}
(G^{1})_{ij} = \left \{
\begin{array}{cl}
V - \{v_{j}\}, & (S)_{ij} \neq \emptyset ~ \wedge ~ i \neq j \\
\emptyset, & (S)_{ij} = \emptyset ~ \vee ~ i = j \\
\end{array},
\right.
\end{equation}
- where $S$ is the walk generator. Walk $G$-coloring is defined with the following iterations
\begin{equation}
\label{e:dc-fw} 
(G^{k+1})_{ij} = \left \{ \begin{array}{cl}
\bigcup_{\beta=1}^{n} (G^{k})_{i\beta} \cap (G^{1})_{\beta j},& i \neq j \\
\emptyset, & i = j \\
\end{array}, \right. ~ k \geq 1
\end{equation}

\section{Cycle problem}
\label{s:cp}

Let us define a $H$-coloring. Arc $H$-coloring is as follows:
\begin{equation}
\label{e:cp-sa}
(H^{1})_{ij} = \left \{
\begin{array}{cl}
(S)_{ij}  & i = j \\
\emptyset, & i \neq j \\
\end{array},
\right.
\end{equation}
- where $S$ is walk generator \ref{e:wg}; and walk $H$-coloring is as follows:
\begin{equation}
\label{e:cp-sw} 
(H^{k+1})_{ij} = \left \{ \begin{array}{cl}
\bigcup_{\beta=1}^{n} (F^{k})_{i\beta} \cap (G^{1})_{\beta j},& i = j \\
\emptyset, & i \neq j \\
\end{array}, \right. ~ k \geq 1,
\end{equation}
- where set matrices $F$ and $G$ were defined in section \ref{s:dc}.
\begin{xca}
\label{x:cp}
For the graph from exercise \ref{x:dm-x2}:
\[
G^{1} = \left (
\begin{array}{cccc}
\emptyset&\{v_{1},v_{3},v_{4}\}&\emptyset&\emptyset \\
\{v_{2},v_{3},v_{4}\}&\emptyset&\{v_{2},v_{3},v_{4}\}&\emptyset \\
\emptyset&\emptyset&\emptyset&\emptyset \\
\emptyset&\emptyset&\emptyset&\emptyset \\
\end{array} \right );
\]
Then,
\[
H^{1} = \left (
\begin{array}{cccc}
\{a_{11}\}&\emptyset&\emptyset&\emptyset \\
\emptyset&\{a_{22}\}&\emptyset&\emptyset \\
\emptyset&\emptyset&\emptyset&\emptyset \\
\emptyset&\emptyset&\emptyset&\emptyset \\
\end{array} \right );
\]
\[
H^{2} = \left (
\begin{array}{cccc}
\{v_{2},v_{3},v_{4}\}&\emptyset&\emptyset&\emptyset \\
\emptyset&\{v_{1},v_{3},v_{4}\}&\emptyset&\emptyset \\
\emptyset&\emptyset&\emptyset&\emptyset \\
\emptyset&\emptyset&\emptyset&\emptyset \\
\end{array} \right ); ~
H^{k} = (\emptyset)_{4 \times 4}, ~ k \geq 3.
\]
Let's notice that non-empty elements of the coloring indicate all cycles in the digraph: loops $a_{11}$, $a_{22}$ and 2-cycle $(v_{1}a_{12}v_{2}a_{21}v_{1})$.
\end{xca}

\begin{lemma}
\label{t:cp1}
In multi digraph $g$, let $p$ be a path of length $k$:
\[
p = (v_{\mu_{1}}a_{\mu_{1}\mu_{2}}v_{\mu_{2}}a_{\mu_{2}\mu_{3}} \ldots v_{\mu_{k}}a_{\mu_{k}\mu_{k+1}}v_{\mu_{k+1}}).
\]
Then,
\[
v_{\mu_{k+1}} \in (F^{1})_{\mu_{1}\mu_{2}} \cap (F^{1})_{\mu_{2}\mu_{3}} \cap \ldots \cap (F^{1})_{\mu_{k}\mu_{k+1}} \subseteq (F^{k})_{\mu_{1}\mu_{k+1}}.
\]
\end{lemma}
\begin{proof}
See proof of lemma \ref{t:dc-s} and decomposition \ref{e:dc-sw-decomposition}.
\end{proof}

\begin{theorem}
\label{t:cp}
In multi digraph $g$, there is a cycle of length $k \geq 1$ attached to vertex $v_{i}$ iff $(H^{k})_{ii} \neq \emptyset$.
\end{theorem}
\begin{proof}
For $k = 1$, the theorem is true due to definition \ref{e:cp-sa}. Let $k > 1$.
\newline\indent
Necessity. Let $c$ be a cycle of length $k$ attached to vertex $v_{i}$:
\[
c = (v_{i}a_{i\mu_{1}}v_{\mu_{1}}a_{\mu_{1}\mu_{2}} \ldots v_{\mu_{k-1}}a_{\mu_{k-1}i}v_{i}),
\]
- where $a_{xy} \in (S)_{xy}$ are arcs and $S$ is the walk generator. Then the following walk
\[
(v_{i}a_{i\mu_{1}}v_{\mu_{1}}a_{\mu_{1}\mu_{2}} \ldots v_{\mu_{k-1}})
\]
is a path of length $k-1$. Then, due to lemma \ref{t:cp1}, 
\[
v_{\mu_{k-1}} \in (F^{k-1})_{i\mu_{k-1}} \neq \emptyset.
\]
On the other hand, due to definition \ref{e:dc-fa}, 
\[
v_{\mu_{k-1}} \in (G^{1})_{\mu_{k-1}i} \neq \emptyset.
\]
Thus, due to definition \ref{e:cp-sw},
\[
v_{\mu_{k-1}} \in (F^{k-1})_{i\mu_{k-1}} \cap (G^{1})_{\mu_{k-1}i} \subseteq (H^{k})_{ii} \neq \emptyset.
\]
\indent
Sufficiency. Let 
\[
(H^{k})_{ii} \neq \emptyset.
\]
Then, due to definition \ref{e:cp-sw}, there exists such couple of sets $(F^{k-1})_{ix}$ and $(G^{1})_{xi}$ that
\[
(F^{k-1})_{ix} \cap (G^{1})_{xi} \neq \emptyset.
\]
Then, due to theorem \ref{t:dc-s0} and definition \ref{e:dc-fa}, there exist a path of length $k-1$ from $v_{i}$ into $v_{x}$ and an arc from $v_{x}$ into $v_{i}$. These path and arc constitute a cycle of length $k$. The cycle is attached to vertex $v_{i}$.
\end{proof}
Let's estimate computational complexity of theorem \ref{t:cp}. By definitions \ref{e:dc-fa}, \ref{e:cp-sw} and due to inclusion \ref{e:dc-v}, 
\[
(H^{k})_{ij} \subseteq V.
\]
Thus, arc $H$-coloring \ref{e:cp-sa} requires $O(n^{2})$ operations. And walk $H$-coloring requires $O(kn^{4})$ operations: $O(kn^{4})$ operations to calculate $F^{k-1}$; $O(n^{3})$ operations to calculate $G^{1}$; and $O(n^{2})$ operations at most to calculate each of $n$ diagonal elements of $H^{k}$. The total computational complexity is 
\[
O(kn^{4}).
\]
Particularly, when $k = n$, theorem \ref{t:cp} solves the Hamilton cycle problem in time $O(n^{5})$. But a simplification is possible. In this case, there is no need for the whole matrix $F^{n-1}$ but only for any one of its strings. That allows solving the Hamiltonian cycle problem with theorem \ref{t:cp} in time 
\[
O(n^{4}).
\]
Obviously, $F$-coloring and $G$-coloring may be swapped in theorem \ref{t:cp}.

\section*{Conclusion}
\label{s:c}

The following polynomial time algorithm detects all paths and cycles of all lengths in form of vertex couples (start, finish):
\begin{description}
\item[Step 1]
Calculate set matrices $F^{1}$, $G^{1}$, and $H^{1}$. Non-empty elements of matrices $F^{1}$ and $H^{1}$ indicate 1-paths and 1-cycles appropriately;
\item[Steps 2 $\div$ n]
Calculate set matrices $F^{k}$ and $H^{k}, ~ k = 1,2,\ldots, n$. Non-empty elements of matrices $F^{k}$ and $H^{k}$ indicate $k$-paths and $k$-cycles appropriately.
\end{description}
\indent
The algorithm could be simplified, for example, with vertex aggregation. In the formulas above, the aggregation might be presented with appropriate box set matrices. The modification may be seen as an incorporation of the DJP algorithm's idea.
\newline\indent
Further time-simplifications could be achieved by mixing the described ``unvisited vertices'' walk coloring with a ``visited vertices'' walk coloring.
\newline\indent
Let us emphasize that the algorithm solves a decision problem ``exist/not exist''. Although, lemma \ref{t:cp1} or a ``visited vertices'' walk coloring can be used for restoration/selection of particular paths/cycles.


\end{document}